\documentclass[review,number,sort&compress]{elsarticle}

\usepackage{lineno}

 \usepackage{graphicx}

\usepackage{amssymb}

\usepackage{color}
\newcommand{\Et}{E_{\rm total}}
\newcommand{\EI}{E_{\rm 1st}}
\newcommand{\EII}{E_{\rm 2nd}}
\newcommand{\EIII}{E_{\rm 3rd}}

\journal{Nuclear Instruments and Methods in Physics Research A}

\begin{document}

\begin{frontmatter}



\title{Reactor antineutrino monitoring
with a plastic scintillator array
as a new safeguards method}

\author[KEK]{S.~Oguri}
\author[minowa]{Y.~Kuroda}
\author[minowa]{Y.~Kato}
\author[minowa]{R.~Nakata}
\author[icepp]{Y.~Inoue}
\author[JAEA]{C.~Ito}
\author[minowa]{M.~Minowa\corref{cor1}}
\cortext[cor1]{Corresponding author}
\ead{minowa@phys.s.u-tokyo.ac.jp}

\address[minowa]{Department of Physics, School of Science, the University of Tokyo, 7-3-1, Hongo, Bunkyo-ku, Tokyo 133-0033, Japan}
\address[KEK]{Institute of Particle and Nuclear Studies, High Energy Accelerator Research Organization (KEK),
1-1, Oho, Tsukuba, Ibaraki 305-0801, Japan}
\address[icepp]{International Center for Elementary Particle Physics, the University of Tokyo,
                 7-3-1, Hongo, Bunkyo-ku, Tokyo 133-0033, Japan}
\address[JAEA]{Oarai Research and Development Center, Japan Atomic Energy Agency,
4002, Naritacho, Oarai-machi, Higashiibaraki-gun, Ibaraki 311-1393,
Japan}

\begin{abstract}
We developed a segmented reactor-antineutrino detector 
made of plastic scintillators for application as a tool in nuclear safeguards inspection
and performed mostly unmanned field operations at a commercial power plant reactor.
At a position outside the reactor building, we measured the difference in reactor antineutrino flux above the ground when the reactor was active and inactive.
\end{abstract}


\end{frontmatter}


\section{Introduction}
A half a century ago, neutrinos were first discovered at a nuclear reactor plant
by Fred Reines, Clyde Cowan and their colleagues%
\cite{Reines-Cowan}.
Indeed, nuclear reactors are the most intense man-controlled sources of neutrinos.
A total flux of 2 $ \times 10^{20}$ antineutrinos/s is emitted by a 1-GW$_{\rm th}$ power plant\cite{flux1,flux2,flux3}.
In recent years, neutrino physics is studied intensively as a means to monitor reactor operations.

International Atomic Energy Agency(IAEA) uses
an extensive set of technical measures
by which it independently verifies
the correctness and completeness
of declarations made by countries about
their stores of nuclear material and activities.
IAEA recommends\cite{iaea_2008_10} 
near-field antineutrino monitoring capabilities
to provide operational status, 
thermal power, and fissile content of reactors
to ensure the implementation of reactor safeguards.
The merits of using antineutrinos physics are as follows:
\begin{itemize}
 \item Non-intrusiveness\\
       Because of their high penetration, antineutrinos
       can be detected outside reactor buildings.
 \item No other sources\\
       Because comparable fluxes of antineutrinos are difficult to create
       without using reactors or accelerators,
       one can therefore obtain raw data of a reactor.
 \item Information of the isotopic content\\
       By measurement of the antineutrino energy spectrum,
       one can determine not only the operational status and
       thermal power of a reactor, but also fissile content\cite{SONGS}.
\end{itemize}

IAEA proposed the development of a compact detector
within a standard 12-meter ISO container
(approximately 25,000 kg net load)
and the aboveground deployment
as medium term (5--8 year timeframe) goals%
\cite{iaea_2008_10}.
The aboveground deployment is very important
because of its non-intrusiveness.
However, it is a challenge because of background noise
induced by cosmic rays
and as yet no group has succeeded in producing a working prototype,
despite several endeavors among various groups
\cite{SONGS, Short_Baseline, Joyo, DANSSino, Nucifer}
around the world.

Taking the above points into account,
we proposed a segmented antineutrino detector, PANDA,
an acronym for plastic anti-neutrino detector array\cite{LesserPANDA}.
Because of its segmented structure and its use of event topology information, PANDA has a strong background rejection capability.

In the first stage of the PANDA project,
we built and operated a small prototype detector
called Lesser PANDA,
at the Unit 3 reactor of the Hamaoka Nuclear Power Plant
of the Chubu Electric Power Co., Inc.
It consisted of 16 modules of plastic scintillators
and a total target mass of 160 kg.
We had planned to measure the change in antineutrino flux
during the startup of the Unit 3 reactor,
but the reactor was not brought online
because of the 2011 Tohoku earthquake off the Pacific Coast of Japan.
We measured background data for two months there.
The results were reported in \cite{LesserPANDA}.

In the next step of our project,
we constructed a 360-kg prototype neutrino detector called PANDA36
as a tool to inspect and assess safeguards.
Over a two-month period we demonstrated its operation above ground
at 36 meters away from the 3.4 GW$_{\rm th}$ reactor core
of Ohi Power Station of the Kansai Electric Power Co., Inc.
The purpose of the experiment was two-fold:
the detection of the change in antineutrino flux
between online and off-line reactor periods and the analysis of background flux
from above-ground measurements for feasibility assessment of project goals.
In this paper, we report results of the experiment using the prototype detector PANDA36.

\section{The detector}
\label{chap:detector}

\begin{figure}
 \centering
 \includegraphics[width=12cm]{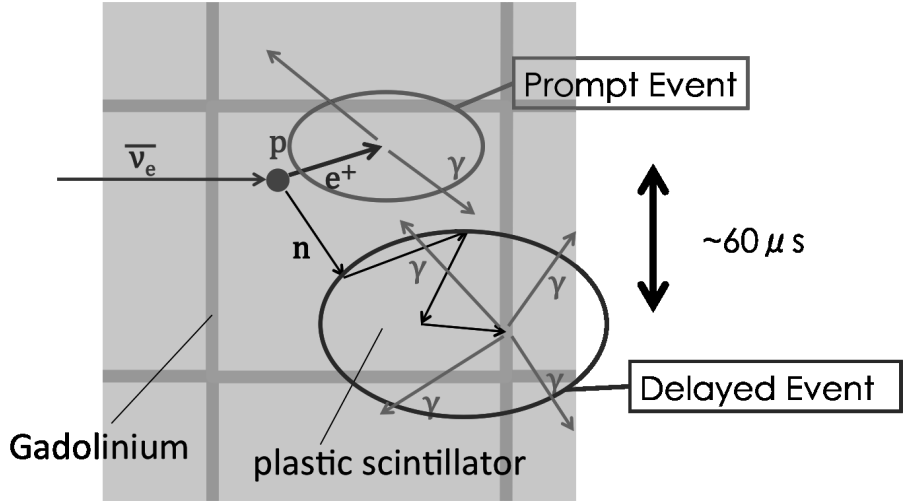}
 \caption{Principle of antineutrino detection
 with a cross-sectional view of the pillar modules}
 \label{fig:method_PANDA_principle}
\end{figure}

\subsection{Principle and features}

We detect antineutrinos via the inverse beta decay interaction on a proton in the plastic
scintillator with the energy threshold of 1.8\ MeV.
\begin{equation}
 \bar{\nu_e} + p \to e^+ + n.
\end{equation}

The positron and the neutron
which are produced by the inverse beta decay
are detected independently.
The positron deposits energy via ionization,
and emits two gamma rays by annihilation:
\begin{equation}
 e^+ + e^- \rightarrow 2 \gamma.
\end{equation}
It is referred to as the prompt event in this paper hereafter.
The neutron is thermalized in the plastic and
captured some time later by gadolinium embedded in between plastic scintillators,
and a gamma ray cascade is produced with total energy of about 8\, MeV:
\begin{equation}
 n + {}^{155}{\rm Gd}
  \rightarrow {}^{156}{\rm Gd}^*
  \rightarrow {}^{156}{\rm Gd} + \gamma\textrm{'s},
\end{equation}
\begin{equation}
 n + {}^{157}{\rm Gd}
  \rightarrow {}^{158}{\rm Gd}^*
  \rightarrow {}^{158}{\rm Gd} + \gamma\textrm{'s}.
\end{equation}
It is referred to as the delayed event.
The prompt and the delayed events are detected
in delayed coincidence.
The principle of detection is illustrated in
Fig.~\ref{fig:method_PANDA_principle}.

Our detector has four original features as follows.

\begin{itemize}
 \item Mobility\\
The target mass of PANDA36 detector is 360kg.
It is small as a neutrino detector.
In addition,
our detector is loaded into a van,
and can operate in the van.
 \item Solid state\\
There are two merits of the solid state scintillator.
The first is the easiness of transportation
compared to liquid scintillator.
Our detector is fully prepared in the van for the measurement
and can be carried to the reactor as it is.
The second is the non-flammability.
Oil-based liquid scintillator is flammable
and flammable oil is prohibited in many cases to be brought to commercial reactor sites.
 \item Aboveground measurement\\
Sea-level operation of reactor neutrino detectors
is one of the greatest issues for the safeguards application.
The previous experiments to detect reactor neutrinos
are conducted in underground sites\cite{Short_Baseline,SONGS}.
In contrast,
our detector is deployed just outside a reactor building.
If we detect the reactor neutrino,
PANDA project is to be the first successful
aboveground experiment.

But it is a difficult challenge.
Above ground,
there are higher background
resulted from cosmic rays.
Especially,
neutrons produced by cosmic muons
are difficult to be discriminated
from delayed events of inverse beta decays.
 \item Segmented detector\\
In order to operate the reactor monitor aboveground,
a powerful background rejection technique is needed.
Our detector is segmented and
the energy deposit in each module is recorded.
So, it becomes possible to use the event topology information
to tag antineutrino events
and to discriminate them from background.

Generally,
liquid scintillators are used for the reactor neutrino experiment
because they are easy to be doped with gadolinium.
But the technique to create clear and colorless
Gd doped plastic scintillator is less established.
Our solution to the issue is to use
the segmented pillar plastic scintillators
which are wrapped in gadolinium coated sheets.
\end{itemize}

\subsection{PANDA36 detector}
\label{sec:detector_PANDA36}

\begin{figure}
 \includegraphics[width=12cm]{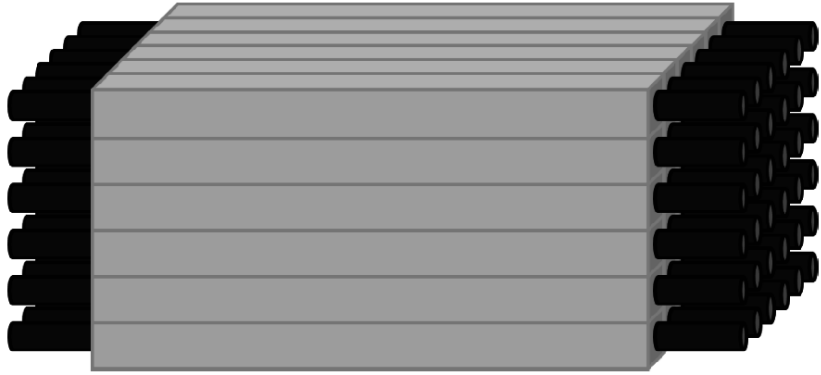}
 \caption{Schematic view of PANDA36 detector}
 \label{fig:detector_PANDA36}
\end{figure}

\begin{figure}
 \includegraphics[width=11cm]{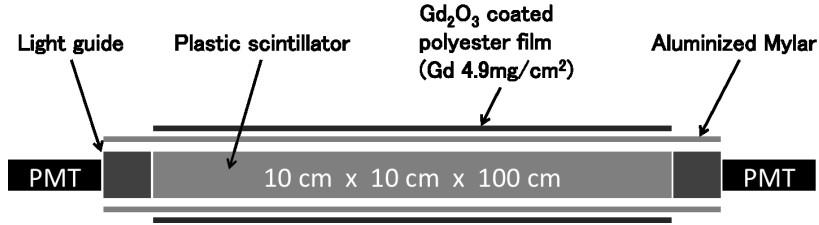}
 \caption{Schematic view of PANDA module}
 \label{fig:detector_PANDAmodule}
\end{figure}

The sketch of PANDA36 detector is shown in
Fig.~\ref{fig:detector_PANDA36}.
The detector consists of 36 identical modules
 which are the same ones as were used in the Lesser PANDA detector\cite{LesserPANDA}.
 The modules are referred to as the PANDA modules.
 Schematic view of a PANDA module is shown Fig.~\ref{fig:detector_PANDAmodule}.
 Each PANDA module has 10\,kg of plastic scintillator
 (EJ-200, ELJEN Technology or RP-408, Rexon Technology) in it.
 Two $10\,{\rm cm}\times10\,{\rm cm}\times10\,{\rm cm}$ acrylic cubic
 light guides are glued to both ends of the plastic scintillator
 with optical cement (EJ-500, ELJEN Technology).
 
Two of 2-inch diameter PMTs(H6410, Hamamatsu)
are glued to the light guides.
The plastic scintillator and the light guides are
wrapped in aluminized Mylar and
gadolinium-oxide coated polyester sheet.
The polyester sheet is obtained
from Ask Sanshin Engineering Corp., Ltd.
The sheet is made of 50 $\mu$m thick polyester film
sandwiched in two layers of 25-$\mu$m thick Gd$_2$O$_3$ coating.
The sheet contains 4.9 mg/cm$^2$ of gadolinium.

\begin{figure}
 \includegraphics[width=12cm]{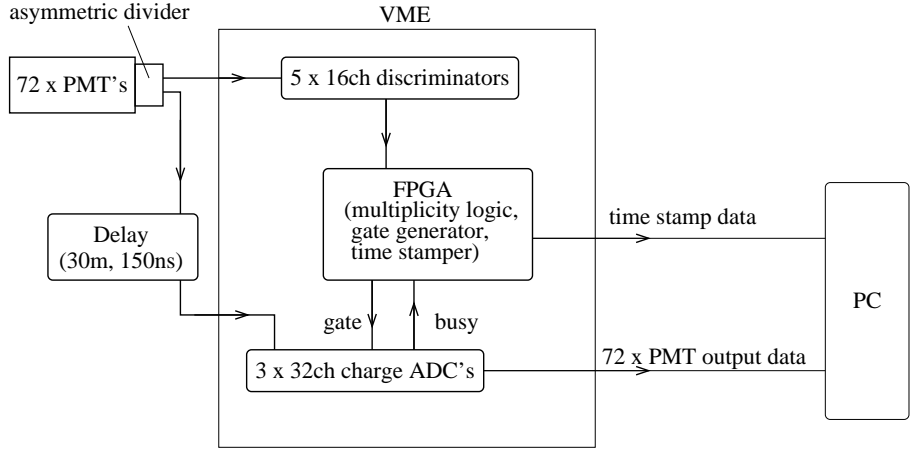}
 \caption{Schematic view of the DAQ system}
 \label{fig:detector_daq}
\end{figure}

\begin{table}
 \centering
 \caption{Table of model numbers of DAQ components }
 \vspace*{3mm}
 \label{tab:detector_daqmodel}
 \begin{tabular}{|c|c|}
  \hline
  component & model number \\
  \hline \hline
  photomultiplier tube & Hamamatsu Photonics H6410 \\
  \hline
  high voltage power supply & Matsusada Precison HARb-2N150 \\
  \hline
  16ch discriminator & CAEN V895 \\
  \hline
  general purpose board (FPGA) & CAEN V1495 \\
  & A395A, A395C \\
  \hline
  32ch charge ADC(QDC)  & CAEN V792 \\
  \hline
 \end{tabular}
\end{table}

Schematic diagram of the DAQ system is shown in
Fig.~\ref{fig:detector_daq}.
The model number of each component
which constructs the DAQ system is
listed in Tab.~\ref{tab:detector_daqmodel}.
The current pulses of PMTs are divided into two lines
by passive signal dividers, where 15\% of
 each current pulse enters a multievent charge ADC (QDC) 
 through a 30 m delay cable(150 ns delay).
The output of QDCs is recorded by a personal computer
via VME bus.
The rest of the current enters
a discriminator and
is then sent to a programmable FPGA board.
The threshold of each discriminator is set to
150 keV energy deposit equivalent
at the far end from the corresponding PMTs.

The FPGA takes coincidence of two PMT signals of each PANDA module
individually and gets 36 module-wise coincidence signals.
The FPGA sends a gate signal of 400-ns width to the QDCs
when there are at least 2 coincidence signals
out of 16 (4 by 4) inner modules.
We did not use 20 outer modules for the event trigger.
At the same time, the FPGA records the time stamp of the leading edge
of the QDC gate.
The time stamps of both the leading and trailing edges of the busy signal
from the QDCs are also recorded when its state has changed.
The time stamp data are stored temporally in an internal FIFO
of the FPGA and are read and recorded by the same PC via VME bus.

Data from the QDCs and the FPGA are 
combined based on the common event numbers
which are embedded in the data.
The event number is incremented by one
on every event independently.

\begin{figure}
 \centering
 \hspace*{-1cm}
 \includegraphics[width=14cm]{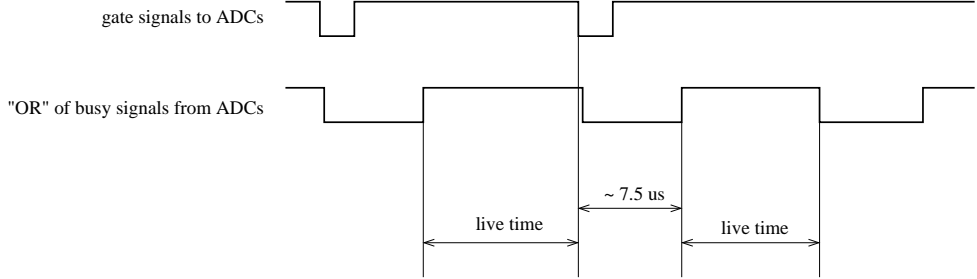}
 \caption{Timing diagram for live times}
 \label{fig:detector_livetime}
\end{figure}

We measured live time of the DAQ
using the time stamp data from FPGA.
In many cases
a busy signal is generated
soon after a gate signal,
and has a width of QDC conversion time of about 7.5 $\mu$s.
The busy signals are also asserted without any
 relating gate signals when the QDCs are waiting for the data transfer.
We took time intervals for the live time while neither the gate
signal to the QDCs nor any of the busy signals of the QDCs were
asserted.
The timing diagram is shown in Fig.~\ref{fig:detector_livetime}.

Energy calibrations with $^{60}$Co gamma ray sources
were carried out on 11th November 2011
at Hongo Campus of the University of Tokyo.
First, energy calibration of each module was carried out
using the Compton edge of $^{60}$Co
because the thickness of each plastic scintillator module
is not enough for the total absorption.

In the next step, the total energy deposit for the detector as a whole
was further calibrated using the total absorption peak of $^{60}$Co
by applying an overall normalization factor to all the modules
since calibration by the Compton edge is prone to be affected by
the uncertainty in the detector model.

The calibration source was inserted from the side
into 1-cm-gap slits between the modules.
Three calibration data at center and both ends
were taken for each module.

There is also time variation of
gains of the PMTs.
We corrected the gains for each data set
using the peak of through-going cosmic muons in the spectrum
of the events in the data set
by the minimum chi square method.
The relativistic muons deposit energy of about 20 MeV
in 10 cm thickness of the plastic scintillator.
The reference data were measured on 11th November 2011
at the same time as the calibration measurement.

\begin{figure}
 \centering
 \includegraphics[width=14cm]{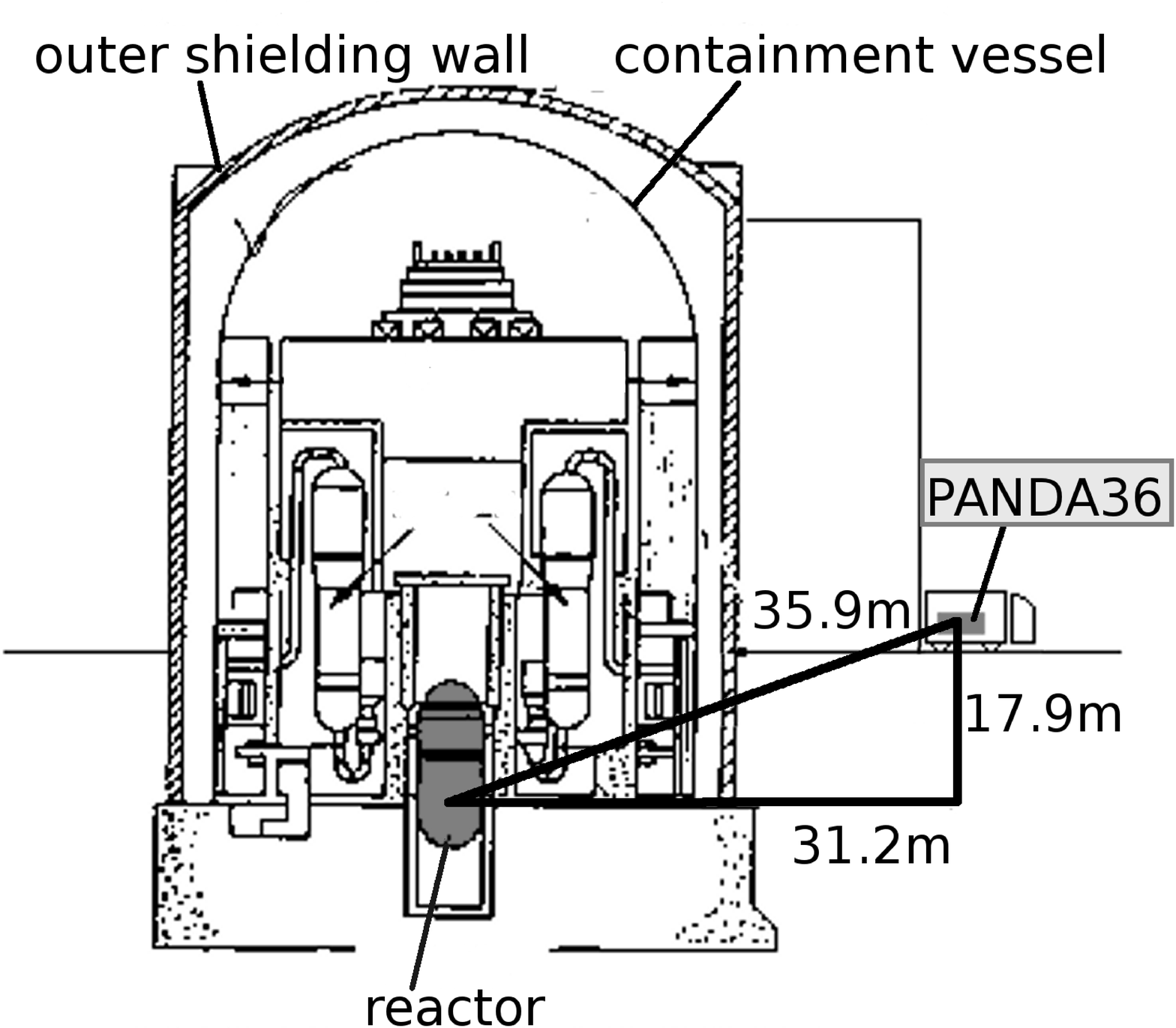}
 \caption{Location of PANDA36 at Ohi Power Station: the van was actually parked in parallel with the building wall.}
 \label{fig:PANDAatOhi}
\end{figure}

PANDA36 was deployed at Ohi Power Station
during the period from 18th November 2011 till 18th January 2012.
The detector in the van was placed at a standoff of 35.9$\pm$0.1\ m from the Ohi Unit 2 reactor core
outside of the reactor building as is illustrated in Fig.~\ref{fig:PANDAatOhi}.
There were neither cosmic ray veto counters nor
passive shields surrounding the detector.
Ohi Unit 2 reactor was in operation at a thermal power
of 3.4 GW$_{\rm th}$.
We continued the measurement even after the reactor shutdown
on 16th December
to take background data at the same place for the rest of the time of about a month.
In the measurement period,
the other reactors(Unit 1, 3, and 4) in Ohi Power Station
were not in operation.

\section{Analysis}
\label{chap:selection}
By the delayed coincidence technique, two kinds of events are picked up
by the data acquisition system.
The first kind is referred to as the correlated event,
in which a prompt event is correlated with a neutron which is captured certain time later by a Gd nucleus.
The second kind is referred to as the uncorrelated event,
which is the accidental coincidence of two independent events caused by natural backgrounds.

Besides antineutrino inverse beta decay interactions,
fast neutrons produced by cosmic ray interactions can form the correlated events.
A fast neutron scatters off a proton and gives a prompt energy deposition,
then is captured by Gd with a characteristic time delay.

There is also another type of correlated background which consists of two
cosmogenic fast neutrons produced at the same time by a muon spallation.
We call it double-neutron correlated background.
If both the neutrons are captured by Gd's and the earlier capture cannot be
discriminated from the positron signal,
they could also cause a correlated background.
This kind of correlated background is, however, 
efficiently eliminated using individual information of the detector segments as is explained below.

Therefore, elimination of the fast neutron correlated background is the key issue
to the detection of relatively small number of antineutrino events 
in overwhelming cosmic ray exposure environment above the ground.
	
We applied the selection cuts shown in
Tab.~\ref{tab:selection_criteria}
to the recorded events to pick up antineutrino events and  to reduce background.

\begin{table}
 \centering
 \caption[Antineutrino and
 fast-neutron event selection criteria]
 {Antineutrino and
 fast-neutron event selection criteria:
 Double underlines denote
 the difference between
 selection 1 and selection 2.}
 \label{tab:selection_criteria}
 \vspace*{2mm}
\begin{tabular}{rcc}
 \hline \hline
 & selection 1 & selection 2 \\
 \hline
 \begin{minipage}[c]{0.22\linewidth}
  \begin{flushright}
   software trigger:
  \end{flushright}
 \end{minipage} &
     \begin{minipage}[c]{0.37\linewidth}
      \centering
      At least two modules in inner 16~modules
      deposite energy of 150~keV or more.
     \end{minipage} &
     \begin{minipage}[c]{0.37\linewidth}
      \centering
      \vspace*{2mm}
      At least two modules in inner 16~modules
      deposite energy of 150~keV or more.
      \vspace*{2mm}
     \end{minipage} \\
 \hline
 \begin{minipage}[c]{0.22\linewidth}
  \begin{flushright}
   prompt:
  \end{flushright}
 \end{minipage} &
     \begin{minipage}[c]{0.37\linewidth}
      \centering
      3~MeV $\le \Et \le$ 6~MeV\\
      \underline{\underline{$\EII \le 520\,{\rm keV}$}}
     \end{minipage} &
     \begin{minipage}[c]{0.37\linewidth}
      \vspace*{2mm}
      \centering
      3~MeV $\le \Et \le$ 6~MeV\\
      \underline{\underline{$\EII \ge 700\,{\rm keV}$}}
      \vspace*{2mm}
     \end{minipage} \\
 \hline
 \begin{minipage}[c]{0.22\linewidth}
  \begin{flushright}
   delayed:
  \end{flushright}
 \end{minipage} &
     \begin{minipage}[c]{0.37\linewidth}
      \centering
      3~MeV $\le \Et \le$ 8~MeV\\
      \vspace*{2mm}
      $\frac{\EIII}{\Et} \ge \frac{\EI / \Et - 0.5}{5}$
     \end{minipage} &
     \begin{minipage}[c]{0.37\linewidth}
      \vspace*{2mm}
      \centering
      3~MeV $\le \Et \le$ 8~MeV\\
      \vspace*{2mm}
      $\frac{\EIII}{\Et} \ge \frac{\EI / \Et - 0.5}{5}$
      \vspace*{2mm}
     \end{minipage} \\
 \hline
 \begin{minipage}[c]{0.22\linewidth}
  \begin{flushright}
   time window:
  \end{flushright}
 \end{minipage} &
     \begin{minipage}[c]{0.37\linewidth}
      \centering
      \underline{\underline{
      $8\,{\rm \mu s} \le t \le 150\,{\rm \mu s}$}}
     \end{minipage} &
     \begin{minipage}[c]{0.37\linewidth}
      \vspace*{2mm}
      \centering
      \underline{\underline{
      $8\,{\rm \mu s} \le t \le 50\,{\rm \mu s}$}}
      \vspace*{2mm}
     \end{minipage} \\
 \hline
 \begin{minipage}[c]{0.22\linewidth}
  \begin{flushright}
   fiducial cut:
  \end{flushright}
 \end{minipage} &
     \begin{minipage}[c]{0.37\linewidth}
      \centering
      The highest energy deposit is
      in inner 16 modules
     \end{minipage} &
     \begin{minipage}[c]{0.37\linewidth}
      \vspace*{2mm}
      \centering
      The highest energy deposit is
      in inner 16 modules
      \vspace*{2mm}
     \end{minipage} \\
 \hline
 \begin{minipage}[c]{0.22\linewidth}
  \begin{flushright}
   muon veto:
  \end{flushright}
 \end{minipage} &
     \begin{minipage}[c]{0.37\linewidth}
      \centering
      There is no event with $\Et > 8$ MeV
      within 250~$\mu$s before the delayed event. 
     \end{minipage} &
     \begin{minipage}[c]{0.37\linewidth}
      \vspace*{2mm}
      \centering
      There is no event with $\Et > 8$ MeV
      within 250~$\mu$s before the delayed event. 
      \vspace*{2mm}
     \end{minipage} \\
 \hline
 \end{tabular}
\end{table}

There are two sets of selection criteria,
``selection 1'' and ``selection 2''.
First,
we picked up the antineutrino-like events by selection 1.
But the selected events also contain
a certain fraction of fast neutron background
because those background events
could not be discriminated from antineutrino events
by selection 1.
Therefore, we introduced selection 2 which is sensitive to fast neutron events.

It should be noted that other background, mainly uncorrelated events,
cannot be fully eliminated by either of two selections because of high background rate at the surface.
Therefore, $N_{\rm s1}$ and $N_{\rm s2}$, the number of events 
by selections 1 and 2, respectively can be written as
\begin{equation}
N_{\rm s1} = \epsilon_{\nu, {\rm s1}} N_\nu + \epsilon_{n, {\rm s1}} N_{n} + \epsilon_{\rm B,s1} N_{\rm B},
\end{equation}
\begin{equation}
N_{\rm s2} = \epsilon_{\nu, {\rm s2}} N_\nu + \epsilon_{n, {\rm s2}} N_{n} + \epsilon_{\rm B,s2} N_{\rm B},
\end{equation}
where $N_\nu$, $ N_{n}$ and $N_{\rm B}$ are numbers of 
antineutrino events, fast neutron events and other background events
occurred in the detector, respectively.
Coefficients
$\epsilon_{\nu, {\rm s1}}$, $\epsilon_{n, {\rm s1}}$, $\epsilon_{\rm B,s1}$,
$\epsilon_{\nu, {\rm s2}}$, $\epsilon_{n, {\rm s2}}$, and $\epsilon_{\rm B,s2}$ are the detection efficiencies of selections 1 and 2
for antineutrinos, fast neutrons and other background events.

From $N_{\rm s1}$ we would like to evaluate the number of fast-neutron-free events $N_{\nu{\rm B}}$
by
\begin{eqnarray}
\label{eq:definition}
N_{\nu{\rm B}} & \equiv  &
N_{\rm s1} - \frac{\epsilon_{n, {\rm s1}}}{\epsilon_{n, {\rm s2}}}N_{\rm s2}\\
& = & 
\left(\epsilon_{\nu, {\rm s1}} -  \epsilon_{\nu, {\rm s2}}\frac{\epsilon_{n, {\rm s1}}}{\epsilon_{n, {\rm s2}}}\right)N_\nu 
+ \left(\epsilon_{\rm B, s1} - \epsilon_{\rm B, s2}\frac{\epsilon_{n,s1}}{\epsilon_{n, {\rm s2}}}\right)N_{\rm B}\\
& = & 
\label{eq:expectation}
\epsilon_{\nu, {\rm s1}}\left(1 -  
\frac{\epsilon_{\nu, {\rm s2}}/\epsilon_{\nu, {\rm s1}}}{\epsilon_{n, {\rm s2}}/\epsilon_{n, {\rm s1}}}\right)N_\nu
+ \left(\epsilon_{\rm B, s1} - \epsilon_{\rm B, s2}\frac{\epsilon_{n,s1}}{\epsilon_{n, {\rm s2}}}\right)N_{\rm B}
\end{eqnarray}

Consequently, $N_{\nu{\rm B}}$ should consist of antineutrino events and 
unnormalized uncorrelated background events,
and is free from fast neutron events.

Detection efficiencies, 
$\epsilon_{\nu, {\rm s1}}, \epsilon_{n, {\rm s1}}, 
\epsilon_{\nu, {\rm s2}}$ and $\epsilon_{n, {\rm s2}}$, are
estimated by Monte Carlo simulation using Geant4 toolkit\cite{Geant4}.
Detection efficiency of selection 1
and the systematic errors are summarized
in Tab.~\ref{tab:efficiency_result}.
A summary of the efficiency ratios, 
$\epsilon_{\nu,{\rm s2}} / \epsilon_{\nu,{\rm s1}}$,
 and
$\epsilon_{n,{\rm s2}} / \epsilon_{n,{\rm s1}}$,
and their systematic errors
are shown in Tab.~\ref{tab:result_efficiencyratio}.

\begin{table}
 \centering
 \caption[Summary of the detection efficiency
 and the systematic error]
 {Summary of the detection efficiency
 and the systematic errors of selection 1:
 The detection efficiency of the delayed events
 is estimated for the simulated events
 which satisfied the prompt event selection.
 The detection efficiency of the time window
 is estimated for the simulated events
 which satisfied the prompt and delayed event selection.
 The relative error consists of
 the uncertainties in the simulation models
 ``relative error(model)''
 and the PMT gain factors
 ``relative error(gain)''.
}
 \label{tab:efficiency_result}

\begin{tabular}{ccccc}
 \hline \hline
 & & efficiency & relative     & relative    \\
 & &            & error(model) & error(gain) \\
 \hline
 prompt
 & trigger             & 28.6\% & 12.1\%$^{(1)}$ &   ---  \\
 & $E_{\rm total}$ cut & 44.2\% & 10.5\%         &   ---  \\
 & $E_{\rm 2nd}$ cut   & 82.2\% & 12.1\%$^{(1)}$ &   ---  \\
 & fiducial cut        & 93.5\% &  5.0\%         &   ---  \\
 & total               &  9.7\% & 16.8\%         &  3.4\% \\
 \hline
 delayed
 & trigger             & 48.8\% & 19.4\%$^{(2)}$ &   ---  \\
 & $E_{\rm total}$ cut & 79.1\% & 19.4\%$^{(2)}$ &   ---  \\
 & $E_{\rm 3rd}$ cut   & 91.9\% & 19.4\%$^{(2)}$ &   ---  \\
 & total               & 35.5\% & 19.4\%         &  1.0\% \\
 \hline
 time window
 &                     & 91.2\% & 14.3\%        & --- \\
 \hline
 Total
 &  $\epsilon_{\rm \nu,s1}=$ & 3.15\% & \multicolumn{2}{c}{$\pm$ 29.6\%} \\
 \hline \\[-4mm]
 \multicolumn{5}{l}{
 \begin{minipage}[t]{0.05\linewidth}
  {\tiny (1)}
 \end{minipage}
 \begin{minipage}[t]{0.9\linewidth}
  The uncertainties
  in the software trigger efficiency
  and the prompt $\EII$ selection
  are estimated as a whole.
 \end{minipage}}\\[5mm]
 \multicolumn{5}{l}{
 \begin{minipage}[t]{0.05\linewidth}
 {\tiny  (2)}
 \end{minipage}
 \begin{minipage}[t]{0.9\linewidth}
  The uncertainties in the three criteria for the delayed events
  are estimated as a whole.
 \end{minipage}}
\end{tabular}
\end{table}
\begin{table}
 \caption[Summary of the efficiency ratios and the systematic errors]
 {Summary of the efficiency ratios and the systematic errors}
 \label{tab:result_efficiencyratio}
\begin{tabular}{rlcc}
 \hline \hline
 & & antineutrino($\epsilon_{\rm \nu,s2} / \epsilon_{\rm \nu,s1}$)
 & fast neutron($\epsilon_{\rm n,s2} / \epsilon_{\rm n,s1}$)\\
 \hline
 relative error: & $\EII$ selection(model)
     & 27~\% & 14.4~\% \\
 relative error: & $\EII$ selection(gain)
     & 1.0~\% & 0.4~\% \\
 relative error: & software trigger
     & --- & 33.3~\% \\
 relative error: & time window(model)
     & 1.1~\% & 1.1~\% \\
 \multicolumn{2}{c}{total relative error}
 & 27~\% & 36.3~\% \\
 \hline
 \multicolumn{2}{c}{value}
 & $0.086 \pm 0.023$ & $1.86 \pm 0.68$ \\
 \hline
\end{tabular}
\end{table}


The selection criteria of selections 1 and 2 are optimized as follows.
First of all, a software trigger was applied to the data for the both selections before the analysis.
Because the hardware thresholds are not necessarily the same for all the PMT's,
it is required to apply the software trigger with a common threshold to estimate the appropriate trigger efficiency.

In the next step,
we selected the prompt events by requiring the total energy $\Et$ to be in the range between 3 and 6\ MeV
to reduce the environmental gamma-ray background.
We expect that a prompt event consists of
one positron and two annihilation gamma rays.
In many cases, $\EI$ corresponds to the ionization loss of
the positron and $\EII$ corresponds to
the Compton scattering of one of the annihilation gamma rays.
Here, $\EI$ and $\EII$ are the highest and the second highest energy deposits
among all the modules.
$\EIII$ is also similarly defined as the third highest deposit energy. 
 To include the energy of 511 keV of the annihilation gamma ray,
 $\EII$ was required to be less than 520 keV for selection 1,
 and to be greater than 700 keV for selection 2 to exclude positrons.
 The double-neutron correlated background events were also efficiently
 eliminated by the $\EII$ cut of selection 1.
 It is because the prompt event of the double-neutron correlated background
 is composed of high energy gamma ray cascade.

\begin{figure}
 \centering
\includegraphics[width=10cm]{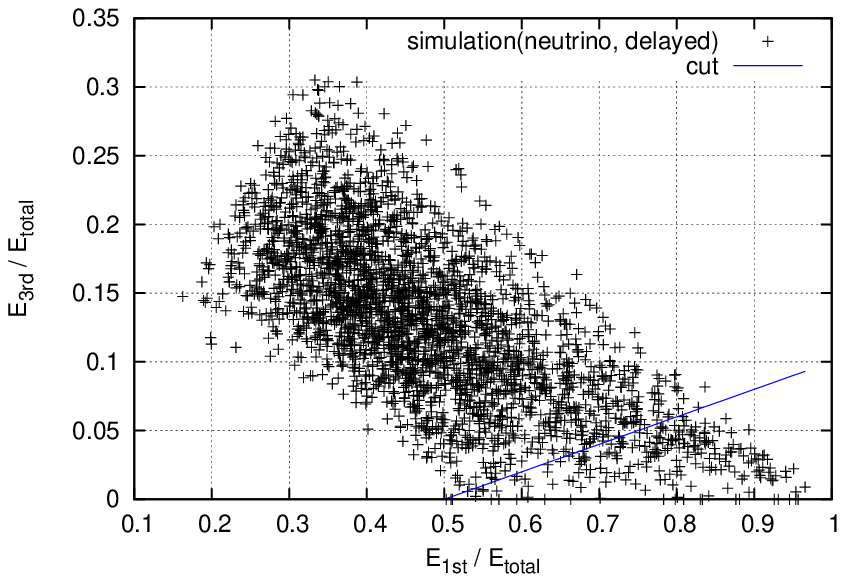}
\includegraphics[width=10cm]{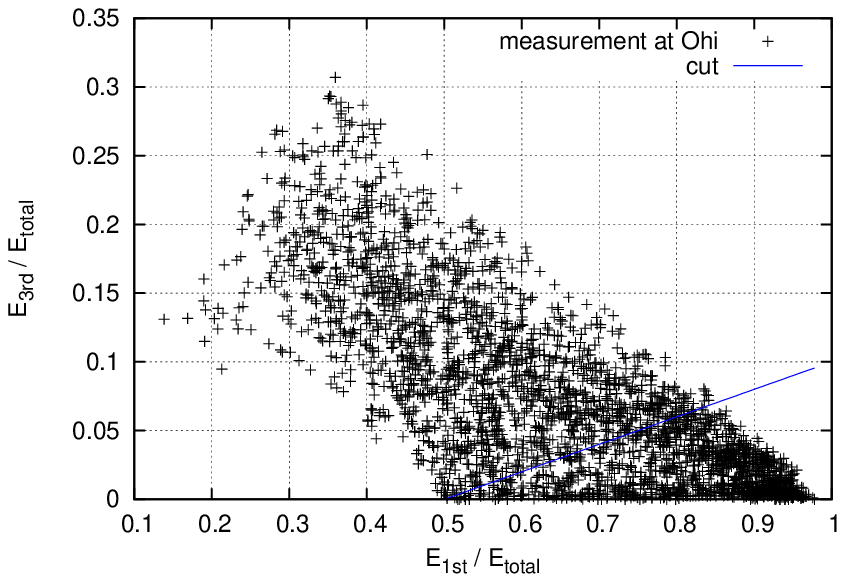}
\vskip 15mm

\caption{
 The above scatter plot shows the simulation
 and the below plot shows the observed data.
 The line shows the selection cut for delayed events
 ($\frac{\EIII}{\Et} \ge\frac{\EI / \Et - 0.5}{5}$). 
 }
\label{fig:selection_delayed_E1E3_scatter}
\end{figure}

The delayed event is characterized by two or more gamma rays as a gamma ray cascade  
with a total energy of 7.9\ MeV
emitted by $^{157}$Gd and 8.5\ MeV
emitted by $^{155}$Gd following the thermal neutron captures common to both the selections.
$\Et$ is, therefore, required to be in the range between 3 and 8\ MeV for the selection of delayed events.
It is rare for the energy deposit to be localized in one module
because two or more gamma rays are emitted at a time
in the delayed event.
So, we expect that
$\EI / \Et$ is not much larger than $\EII / \Et$ nor $\EIII / \Et$.
Fig.~\ref{fig:selection_delayed_E1E3_scatter}
shows scatter plots of $\EIII / \Et$
vs $\EI / \Et$
of the simulation(above)
and the observed data(below).
As expected,
the events of the observed data are
concentrated on $\EI / \Et \sim 1$
because they are dominated by gamma ray background events,
and the events of the simulation
are scattered.
We accordingly required
\begin{equation}
 \frac{\EIII}{\Et} \ge
  \frac{\EI / \Et - 0.5}{5}
\end{equation}
in both selection 1 and selection 2.

We paired a prompt-like event with the following delayed-like event
when the delay time was within a predefined time window.

\begin{figure}
 \centering
 \includegraphics[width=0.7\linewidth]{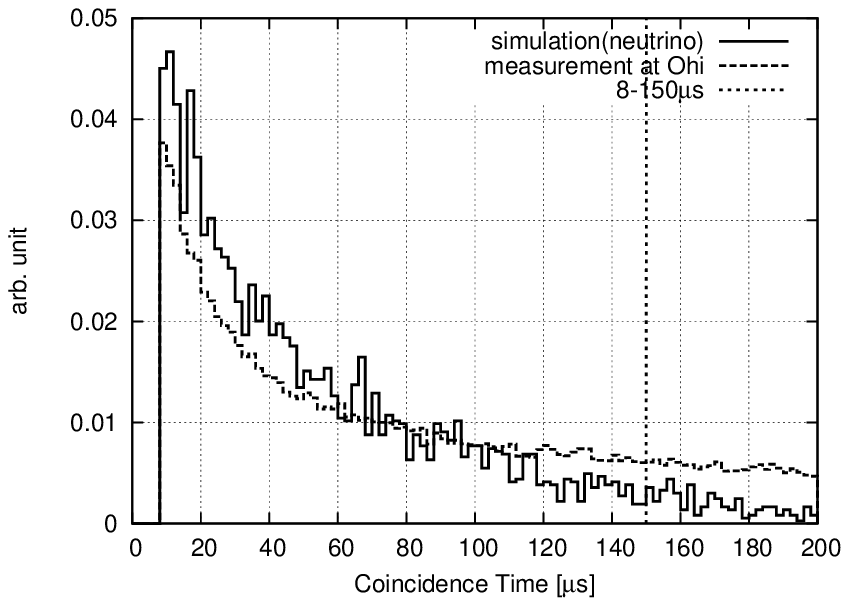}
 \includegraphics[width=0.7\linewidth]{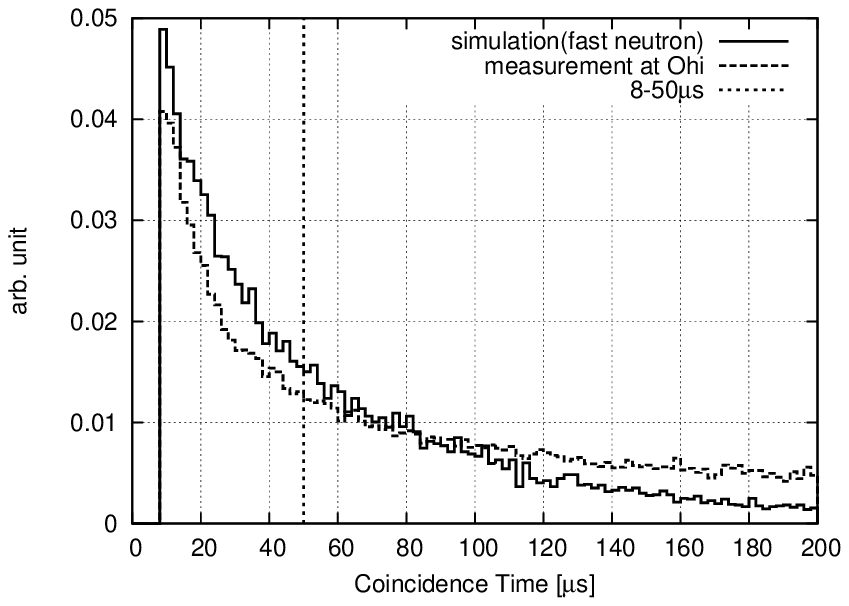}
\vskip 15mm

 \caption[Distribution of the prompt-delayed interval]
 {Distribution of the prompt-delayed interval:
 The prompt events satisfied the software trigger criterion,
 $\Et$ selection and $\EII$ selection.
 The delayed events satisfied the software trigger criterion,
 $\Et$ selection and $\EIII / \Et$ selection.
 (top) The solid and dashed lines
 show the simulated(antineutrino) and observed distributions
 which are selected by selection 1.
 (bottom) The solid and dashed lines
 show the simulated(fast neutron) and observed distributions
 which are selected by selection 2.
 We set the time windows as
 $8\,{\rm \mu s} \le t \le 150\,{\rm \mu s}$(selection 1)
 and $8\,{\rm \mu s} \le t \le 50\,{\rm \mu s}$(selection 2).
 }
 \label{fig:analysis_Time}
\end{figure}

Due to the conversion time of the
QDC(about 7.5 $\mu$s),
the minimum threshold for the time window was
set at 8 $\mu$s.
The distributions of the prompt-delayed intervals
by selection 1(top) and selection 2(bottom)
are shown in Fig.~\ref{fig:analysis_Time}.

The dashed lines show the observed data and the solid lines show the simulation of neutrino events(top)
and fast neutron correlated events(bottom).
The simulation curves for both the selections are exponentially decreasing with time
because the event pair is correlated.
On the other hand, the data curves are decreasing more gradually.
It is most probably due to accidental coincidence events pairs, which distribute constantly over the time.

Therefore,
the shorter the time window is,
the higher the fraction
of the correlated events is selected.
But setting shorter time window leads to
reduction in detection efficiency.
We set the time windows as
\begin{equation}
 8\,{\rm \mu s} \le t \le 150\,{\rm \mu s}
  \quad ({\rm selection\ 1})
\end{equation}
so as to get high efficiency for the low-rate
antineutrino events,
and
\begin{equation}
 8\,{\rm \mu s} \le t \le 50\,{\rm \mu s}
  \quad ({\rm selection\ 2})
\end{equation}
to get high-purity fast neutron events sample
of sufficient rate even with low efficiency.

Next we focus on the position of the highest energy deposit
in the prompt event.
Because $\EI$ of the prompt event is supposed to correspond to
the ionization loss of the positron,
the position of the highest energy deposit is
uniformly distributed
in 36 modules.
On the other hand,
the prompt $\EI$ of the correlated background
by the fast neutron is supposed to correspond to the proton recoil.
Because neutron interaction with a hydrogen nucleus
at the neutron energy of 10 MeV has
a cross section of about 1 barn,
neutrons have the mean free path of about 15 cm
in the plastic scintillator.
Therefore, the highest energy deposit tends to occur in
outer 20 modules.
However, it should be noted that
because of the correlation with
the software trigger criterion,
both the distributions already concentrate on inside 16 modules.
We cut the events whose prompt $\EI$ module are located
in outer 20 modules as a fiducial cut.

We introduce a muon veto cut by software.
Cosmic ray muons can produce
fast neutrons which bring on
correlated background.
We assume the event with $\Et$ of more than 8 MeV
is a muon candidate
and rejected any prompt-delayed event pairs
in which a muon candidate event occurred
within 250 $\mu$s before the delayed event .

The detection efficiency
of selection 1(Tab.~\ref{tab:selection_criteria})
was calculated using the simulation toolkit Geant4.
The systematic uncertainty of the efficiency
was estimated as follows.
We assumed that the systematic uncertainty
consists of two different mechanisms,
uncertainties in the simulation models
and uncertainties in the PMT gain factors.
To estimate the uncertainties in the simulation models,
dedicated experiments with radioactive sources were carried out
and the detection rates were compared
between the observation and the simulation.
To estimate the uncertainty
in the prompt software trigger efficiency
and the prompt $\EII$ selection,
we carried out an experiment
with $^{22}$Na positron- and gamma ray source.
And to estimate the uncertainty
in the selection criteria for delayed events,
we carried out an experiment
with $^{252}$Cf neutron source.
Because it is difficult to verify the prompt $\Et$ selection
by experiments using radioactive sources,
we calculated its uncertainty using the simulation result
and the estimated energy resolution.
To estimate the uncertainty of the detection efficiency
attributed to the PMT gain variation,
we also used the simulation result,
but not experimental data.

The summary of the detection efficiency
and the systematic errors is
shown in Tab.~\ref{tab:efficiency_result}.
The detection efficiencies of the prompt event selection,
the delayed event selection and the time window
are affected by the position of the inverse beta decay.
There are, therefore, correlations among them.
The relative error of 29.6~\% and
the detection efficiency of $(3.15\pm0.93)$~\%
were estimated.

We need to estimate the efficiency ratios
$\epsilon_{\rm \nu,s2} / \epsilon_{\rm \nu,s1}$ and
$\epsilon_{n, \rm s2} / \epsilon_{n, \rm s1}$ 
to calculate $N_{\rm \nu B}$ in Eq.(\ref{eq:definition})
and to evaluate the theoretical expectation by Eq.(\ref{eq:expectation}). 

The ratios of the detection efficiencies
were calculated using the simulation.
The differences between selection 1 and selection 2
are the prompt $\EII$ selection and the coincidence time window.
Contributions of the other common selection criteria
cancel out each other.
Both uncertainties in the simulation models and the PMT gain factors
contribute to the uncertainty in the prompt $\EII$ selection.
On the other hand, only the uncertainties in the simulation models
contribute to the uncertainty of the coincidence time window cut.
To estimate the uncertainties in the simulation models,
we compared the energy spectra and the coincidence time distributions
between the simulation and the observed data.

The expected antineutrino event rate was thus estimated
by the factor of the first term of Eq.(\ref{eq:expectation}).
According to a report released
by the Kansai Electric Power Co., Inc.,
the reactor generated the thermal power of $3.4 \pm 0.1$ GW.
The systematic error of the thermal power is not reported,
so the very conservative value was used for this estimation.
We assumed in this estimation that
the fission fuel fraction is the same as the SONGS experiment\cite{SONGS}.
It is simply assumed that all the fuel is concentrated 
at the center of the reactor core as it is a sufficient approximation\cite{Shimazu}
for the present experiment. 
The expected antineutrino detection rate
by PANDA36(target mass: $360 \pm 18$ kg) is
$17.3\pm 6.2$~events/day.

\section{Result}
\begin{figure}
 \centering
\includegraphics[width=10cm]{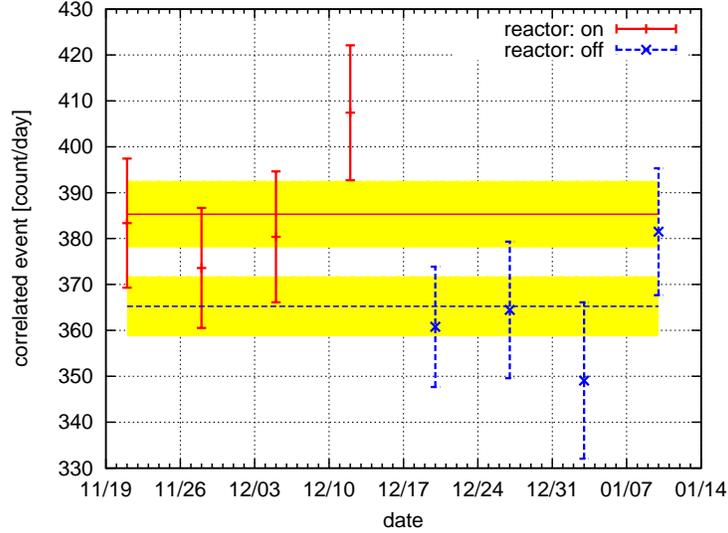}
\vskip 15mm
\caption[$N_{\rm \nu B}$ event rate]
 {Seven day average daily event rates $N_{\rm \nu B}$}
\label{fig:final_number}
\end{figure}

We calculated the daily rates $N_{\rm \nu B}$ by Eq.(\ref{eq:definition})
averaged for seven days 
and plotted them in Fig.~\ref{fig:final_number}.
The reactor went shutdown on 16th December.
The crosses and the stars show
the data during the reactor ON and OFF period, respectively.
The errors shown are statistical only because the systematic error
of $\epsilon_{n,{\rm s1}} / \epsilon_{n,{\rm s2}}$ is attributed to the overall
normalization of $N_{\nu {\rm B}}$.
The horizontal bands represent averages and error interval of 
$N_{\nu {\rm B}}$
for the reactor ON period and OFF period.

The reactor ON/OFF difference of $N_{\rm \nu B}$
was evaluated to be $21.8 \pm 11.4$~events/day.
The result is consistent with the predicted event rate
of $17.3 \pm 6.2$~events/day.

We assumed the correlated background consists of only fast neutron events by now.
We discuss other candidates of long-lived cosmic ray activation products in the following.

Beta decays of $^9$Li and $^8$He produced by cosmic muons could also cause correlated background.
Decay rates of $^9$Li and $^8$He 
are estimated to be less than
$4 \times 10^{-7} \mu^{-1} {\rm g^{-1} cm^2}$ and
$4 \times 10^{-8} \mu^{-1} {\rm g^{-1} cm^2}$,
respectively\cite{PhysRevC.81.025807}.
Assuming the vertical muon intensity of
$1.0 \times 10^{-2}~{\rm cm^{-2}sr^{-1}s^{-1}}$
and angular dependence of $\cos^2 \theta$,
production rates of the isotopes
in PANDA36($3.6 \times 10^5$~g) were estimated to be
\begin{equation}
 {\rm ^9 Li}:~~
3.0 \times 10^{-3}~{\rm s^{-1}} \sim 260~{\rm day^{-1}},
\end{equation}
\begin{equation}
 {\rm ^8 He}:~~
3.0 \times 10^{-4}~{\rm s^{-1}} \sim 26~{\rm day^{-1}}.
\end{equation}
Because the detection efficiency of less than 0.5~\%
was calculated of each beta decay by the simulation
and no ON/OFF difference in the muon flux is expected,
contribution to the final result is negligible.

Natural radioisotopes
$^{220}$Rn, $^{222}$Rn and their daughters
emit $\alpha$ radiation with the energy range of
4--9~MeV(1--2~MeV electron equivalent).
If these isotopes permeate into the plastic scintillator,
they might cause ($\alpha$, n) reactions.
Such events could also be observed as the correlated background
by the delayed coincidence between the $\alpha$ ionization and
the following neutron capture.
However, the abundance of the radon near the detector
in the open air is expected to be
much less than under the ground,
and the plastic scintillator is less permeable by radon
than liquid scintillator.
In addition to that,
no ON/OFF difference is expected
in the event rate from radon and their daughters
because the radon is not generated
by the fission in the reactor.
It should be noted that
the prompt event of the ($\alpha$, n) reaction
is rejected by the $\Et$ cut of selection 1 and selection 2.

\section{Prospect}

We plan to build PANDA100, an antineutrino detector with
$10\times10$ modules, as our ultimate goal
by upgrading PANDA36,
possibly with an intermediate prototype PANDA64
with $8\times 8$ modules.
They are expected to have higher detection efficiency
than PANDA36 in addition to a larger target mass
because the escape of the cascade gamma rays
following neutron capture
is suppressed with the larger volume.

Antineutrino detection efficiency of PANDA100
is estimated to be 9.24~\% using simulation
by applying a selection similar to selection 1.
If PANDA100 were deployed at the same position as PANDA36
at Ohi Power Station,
the selected antineutrino event rate would be
$\sim 147\ {\rm events/day}.$
Background rates of PANDA100 can also be estimated assuming the same background fast neutron flux as PANDA36.
PANDA100 is thus expected to be able to detect the change
of the reactor status by more than 5$\sigma$ in a week aboveground
and to achieve the IAEA's medium term goals.

In order to make more precise measurement of the antineutrino flux,
we have to reject more background events.
As discussed in the former sections,
the main source of the background is
the ``proton recoil -- neutron capture'' events by fast neutrons.
One of the solutions to the background rejection is
to shield the detector from fast neutrons
with water tanks or polyethylene blocks.
Our detector has the ability to discriminate between
the prompt events and the delayed events.
So if fast neutrons could be thermalized before reaching the detector,
we would be able to reject the background events of fast neutrons.

\section{Conclusion}

We developed the prototype of the reactor antineutrino detector
as a new safeguards tool
and demonstrated the almost unmanned field operation
at the reactor site for two months.
We observed the difference of
the reactor antineutrino flux
with the reactor ON and OFF
even with small 360-kg prototype detector
above the ground in the vicinity of a commercial reactor of a power plant for the first time.

Our detection efficiency of the inverse beta decay
is $3.15 \pm 0.93$~\%.
We installed the detector at $35.9 \pm 0.1$~meters
away from the $3.4 \pm 0.1$~GW$_{\rm th}$ reactor core.
The difference of the antineutrino event rate
between the reactor ON period 
and the reactor OFF period 
is $21.8 \pm 11.4$ events/day.
The predicted difference is $17.3 \pm 6.2$ events/day.

Assuming the fast neutron flux measured by the PANDA36 experiment,
the ultimate 100-module detector, PANDA100, is expected to be able to detect the change
of the reactor status by more than 5$\sigma$ in a week aboveground
and to achieve the IAEA medium term goals.

\section*{Acknowledgements}
The authors thank the Kansai Electric Power Co., Inc. for its cooperation for our experiment on site of Ohi Power Station.
They are also grateful to Professor Yoichi Fujiie for his support for this research project.
The authors wish to acknowledge useful discussions with professor 
Yoichiro Shimazu.
This research was partially supported by the Japanese Ministry of Education, Science, Sports and Culture, Grant-in-Aid for COE Research, Grant-in-Aid for Scientific Research (B), and Grant-in-Aid for JSPS Fellows and also by the Mitsubishi Foundation.




\end{document}